\documentclass[twocolumn]{webofc}
\usepackage[varg]{txfonts}   
\usepackage[singlelinecheck=false]{caption}
\usepackage{subcaption}
\usepackage{xcolor}
\newcommand\ecm{$e~\mathrm{cm}~$}
\newcommand\ecmm{e~\mathrm{cm}}
\newcommand{\corrcol}[1]{\textcolor{black}{#1}}
\begin{document}
\title{The n2EDM experiment at the Paul Scherrer Institute}
%
%

\author{
  \firstname{C.} \lastname{Abel}\inst{1} \and
  \firstname{N. J.} \lastname{Ayres}\inst{1} \and
  \firstname{G.} \lastname{Ban}\inst{2} \and
  \firstname{G.} \lastname{Bison}\inst{3} \and
  \firstname{K.} \lastname{Bodek}\inst{4} \and
  \firstname{V.} \lastname{Bondar}\inst{5} \and
  \firstname{E.} \lastname{Chanel}\inst{6} \and
  \firstname{P.-J.} \lastname{Chiu}\inst{3} \and  
  \firstname{B.} \lastname{Clement}\inst{7} \and
  \firstname{C.} \lastname{Crawford}\inst{8} \and  
  \firstname{M.} \lastname{Daum}\inst{3} \and
  \firstname{S.} \lastname{Emmenegger}\inst{9} \and  
  \firstname{P.} \lastname{Flaux}\inst{2} \and
  \firstname{L.} \lastname{Ferraris-Bouchez}\inst{7} \and
  \firstname{W. C.} \lastname{Griffith}\inst{1} \and
  \firstname{Z. D.} \lastname{Grujić}\inst{10} \and
  \firstname{P. G.} \lastname{Harris}\inst{1} \and
  \firstname{W.} \lastname{Heil}\inst{11} \and
  \firstname{N.} \lastname{Hild}\inst{3} \and
  \firstname{K.} \lastname{Kirch}\inst{3,9} \and
  \firstname{P. A.} \lastname{Koss}\inst{5}\fnsep\thanks{peter.koss@kuleuven.be / ORCiD: 0000-0001-5094-3056} \and
  \firstname{A.} \lastname{Kozela}\inst{12} \and
  \firstname{J.} \lastname{Krempel}\inst{9} \and
  \firstname{B.} \lastname{Lauss}\inst{3}\fnsep \thanks{bernhard.lauss@psi.ch / ORCiD: 0000-0002-1986-391X} \and
  \firstname{T.} \lastname{Lefort}\inst{2} \and
  \firstname{Y.} \lastname{Lemière}\inst{2} \and
  \firstname{A.} \lastname{Leredde}\inst{7} \and
  \firstname{P.} \lastname{Mohanmurthy}\inst{3,9} \and
  \firstname{O.} \lastname{Naviliat-Cuncic}\inst{2}\fnsep\thanks{Present address: Michigan State University, East-Lansing, MI, USA} \and
  \firstname{D.} \lastname{Pais}\inst{3} \and
  \firstname{F. M.} \lastname{Piegsa}\inst{6} \and
  \firstname{G.} \lastname{Pignol}\inst{7} \and
  \firstname{M.} \lastname{Rawlik}\inst{9} \and
  \firstname{D.} \lastname{Rebreyend}\inst{7} \and
  \firstname{D.} \lastname{Ries}\inst{13} \and
  \firstname{S.} \lastname{Roccia}\inst{14} \and
  \firstname{K.} \lastname{Ross}\inst{13} \and  
  \firstname{D.} \lastname{Rozpedzik}\inst{4} \and
  \firstname{P.} \lastname{Schmidt-Wellenburg}\inst{3} \and
  \firstname{A.} \lastname{Schnabel}\inst{15} \and
  \firstname{N.} \lastname{Severijns}\inst{5} \and
  \firstname{J.} \lastname{Thorne}\inst{1,6} \and
  \firstname{R.} \lastname{Virot}\inst{7} \and  
  \firstname{J.} \lastname{Voigt}\inst{15} \and
  \firstname{A.} \lastname{Weis}\inst{10} \and  
  \firstname{E.} \lastname{Wursten}\inst{5} \and  
  \firstname{J.} \lastname{Zejma}\inst{4} \and
  \firstname{G.} \lastname{Zsigmond}\inst{3}
}

\institute{University of Sussex, Brighton, United Kingdom
\and
	Normandie Univ, ENSICAEN, UNICAEN, CNRS/IN2P3, LPC Caen, Caen France
\and
    Paul Scherrer Institute, 5232 Villigen, Switzerland 
\and
    Jagiellonian University, Cracow, Poland
\and
	Instituut voor Kern- en Stralingsfysica, KU Leuven, 3001 Heverlee, Belgium 
\and
	Laboratory for High Energy Physics and Albert Einstein Center for Fundamental Physics, University of Bern, Bern, Switzerland
\and
	Laboratoire de Physique Subatomique et de Cosmologie, Grenoble, France
\and
	University of Kentucky, Lexington, United States of America	
\and
	Institute for Particle Physics and Astrophysics, ETH Zürich, 8093 Zürich, Switzerland
\and
	Physics Department, University of Fribourg, 1700 Fribourg, Switzerland	
\and
	Institut für Physik, Johannes-Gutenberg-Universität, Mainz, Germany
\and
	Henryk Niedwodniczański Institute for Nuclear Physics, Cracow, Poland
\and
	Institut für Kernchemie, Johannes-Gutenberg-Universität, Mainz, Germany
\and
	CSNSM, Universit\'e Paris Sud, CNRS/IN2P3, Universit\'e Paris Saclay, Orsay-Campus, France
\and
	Physikalisch Technische Bundesanstalt, Berlin, Germany
}

\abstract{%
We present the new spectrometer for the neutron electric dipole moment \corrcol{(nEDM)} search at the Paul Scherrer Institute \corrcol{(PSI)}, called n2EDM.
The setup is at room temperature in vacuum using ultracold neutrons.
n2EDM features a large UCN double storage chamber design with neutron transport adapted to the PSI UCN source.
%
%
The design builds on experience gained from the \corrcol{previous} apparatus operated at PSI until 2017.
An order of magnitude increase in sensitivity is calculated for the \corrcol{new} baseline setup based on scalable results from the previous apparatus, and the UCN source performance achieved in 2016.
%
%
}
\maketitle
\section{Introduction}
%
%
%
A static neutron electric dipole moment (nEDM) would violate parity P and time reversal T symmetries.
After the observations of both P and CP violation in weak decays~\cite{wu1957,christenson1964}, one expects a non-zero contribution to the nEDM from the weak sector; here we neglect the possibility of CP violation in the strong sector.
The contribution from the weak sector yields the Standard Model estimate of the size of the nEDM on the order of $10^{-32}$ \ecm~\cite{khriplovich1989}, well below the present best experimental limit of $3 \times 10^{-26}$ \ecm (90\% C.L.) \cite{pendlebury2015}.\\
Thus, nEDM searches integrate well in investigations into new sources of CP-violation in nature~\cite{ramsey1982}, which point towards beyond Standard Model (BSM) physics~\cite{engel2013}.
%
%
A long standing problem only solvable with BSM physics is baryogenesis, namely the production of an imbalance of matter over anti-matter in the early Universe.
The three basic criteria necessary for baryogenesis were formulated by Sakharov~\cite{sakharov1967violation}.
One of them is a new source of CP violation, orders of magnitude larger than in the Standard Model.
Several BSM models incorporate stronger CP violation~\cite{morrissey2012} and at the same time predict much larger EDMs for fundamental particles.
Thus, finding a non-zero nEDM would contribute to the understanding of baryogenesis.\\
The possible existence of static electric dipole moments of fundamental particles was formulated as a question to experimental physics almost 70 years ago~\cite{purcell1950}, followed by the first upper limit of the nEDM in 1957~\cite{smith1957}.
Since then, a long line of experiments have pushed the limit down by six orders of magnitude~\cite{lamoreaux2009}.
These investigations are complementary to many other experimental and theoretical efforts in low- and high-energy physics~\cite{engel2013,chupp2015,schmidt2016,shu2013}.\\
Our collaboration has a staged \corrcol{experimental program}~\cite{baker2011}.
In the initial phase, we made use of an upgraded version of the RAL/Sussex/ILL spectrometer~\cite{baker2014}.
The acquired data~\cite{roccia2018proc} will allow us to improve on the present best upper limit~\cite{pendlebury2015}.
Additionally, during the commissioning and data taking of the initial \corrcol{experimental} phase, our collaboration has been developing a completely new spectrometer called n2EDM.
It is a room temperature in vacuum experiment using ultracold neutrons (UCN).
The spectrometer design combines the pioneering PNPI double chamber design~\cite{altarev1980} and a Hg co-magnetometry system~\cite{green1998}.
It draws from the expertise in technical development and systematics control of our collaboration~\cite{pendlebury2015}.
%
%
%
\section{Experimental method}
All nEDM experiments look for a coupling of the neutron spin to an applied electric field on top of a known magnetic coupling.
This is illustrated by the Hamiltonian of a neutron in a magnetic and electric field
\begin{equation}
  H = -\vec{\mu}_{\rm n} \cdot \vec{B} - \vec{d}_{\rm n} \cdot \vec{E}~,
  \label{eq:hamiltonian}
\end{equation}
where $\vec{\mu}_{\rm n}$ is the magnetic dipole moment, $\vec{d}_{\rm n}$ the electric dipole moment, $\vec{B}$ and $\vec{E}$ are the magnetic and electric fields.\\
Since the first nEDM result in 1957, almost all experiments have been using the Ramsey method of time separated oscillating fields~\cite{ramsey1949}.
With this method measurements of the Larmor precession frequency of polarized neutrons in a magnetic field are performed.
In our apparatus, the Larmor frequency is measured in the two cases of parallel/anti-parallel $\vec{B}$ and $\vec{E}$ fields.
The Larmor frequencies are given following Eq.  (\ref{eq:hamiltonian})
\begin{align}
  h \nu_{\uparrow \uparrow} &= -2(\mu_{\rm n} B_{\uparrow \uparrow} + d_{\rm n} E_{\uparrow \uparrow}) \label{eq:upup}\\
  h \nu_{\uparrow \downarrow} &= -2(\mu_{\rm n} B_{\uparrow \downarrow} - d_{\rm n} E_{\uparrow \downarrow})~\label{eq:updown},
\end{align}
where $\uparrow \uparrow$ stands for parallel $\vec{B}$/$\vec{E}$-fields and $\uparrow \downarrow$ stands for anti-parallel $\vec{B}$/$\vec{E}$-fields.\\
The nEDM, $d_{\rm n}$, can be extracted from a differential measurement between the frequencies $\nu_{\uparrow \uparrow}$ and $\nu_{\uparrow \downarrow}$ with
\begin{equation}
  d_{\rm n} = \frac{h(\nu_{\uparrow \downarrow} - \nu_{\uparrow \uparrow}) - 2 \mu_{\rm n} (B_{\uparrow \uparrow} - B_{\uparrow \downarrow})}{2(E_{\uparrow \uparrow} + E_{\uparrow \downarrow})}~.
\end{equation}
The statistical sensitivity \corrcol{of a single measurement with this method is given by}
\begin{equation}
  \sigma (d_{\rm n}) = \frac{\hbar}{2 \alpha | E | T \sqrt{N}}~,
\label{eq:sens}
\end{equation}
where $| E |$ is the electric field strength, $T$ is the free precession time of the neutrons, $N$ is the number of counted neutrons and $\alpha$ is a measure of the neutron polarization.
The analysis of a single chamber experiment, as with our \corrcol{previous} apparatus, uses corrected neutron Larmor frequencies~\cite{roccia2018proc}.
These frequencies are corrected using $\mathcal{R}$-values, which are given by
\begin{equation}
\mathcal{R} = \frac{\nu_{\rm n}}{\nu_{\rm Hg}} = \bigg| \frac{\gamma_{\rm n}}{\gamma_{\rm Hg}} \bigg| \bigg(1 + \frac{\langle z \rangle G}{B_0}\bigg) + \frac{E}{\pi h \nu_{\rm Hg}} d_{\rm n}~,
\label{eq:singleR}
\end{equation}
where $\gamma_{\rm n}$, $\gamma_{\rm Hg}$ are the gyromagnetic ratios of the neutron (resp. mercury), $G$ is the vertical linear magnetic gradient over the UCN chamber, $E$ is the electric field and $\langle z \rangle$ is the center-of-mass difference between the Hg atoms and the UCN.\\
In the case of a double chamber design one can use the $\mathcal{R}$-values for both chambers in the analysis.
The difference between the top ($\mathcal{R}^{\rm T}$) and the bottom ($\mathcal{R}^{\rm B}$) chamber $\mathcal{R}$-values yields
\begin{equation}
\mathcal{R}^{\rm T} - \mathcal{R}^{\rm B} = \frac{2E}{\pi h \nu_{\rm Hg}} d_{\rm n} + \bigg| \frac{\gamma_{\rm n}}{\gamma_{\rm Hg}} \bigg| (\langle z \rangle ^{\rm T} - \langle z \rangle ^{\rm B}) \frac{G}{B_0}~,
\label{eq:doubleR}
\end{equation}
where $\langle z \rangle ^{\rm T}$, $\langle z \rangle ^{\rm B}$ are the differences in center-of-mass between the Hg atoms and the UCN in the top (resp. bottom) chamber.
From eq. (\ref{eq:doubleR}) we immediately see that the gradient induced systematic is strongly suppressed in a double chamber setup \corrcol{to a contribution proportional only to the difference in center-of-mass, i.e., $\langle z \rangle ^{\rm T} - \langle z \rangle ^{\rm B}$.}
%
%
\section{n2EDM apparatus}
\label{sec:n2EDM}
\begin{figure}[h]
\centering
  \includegraphics[width=0.5\textwidth]{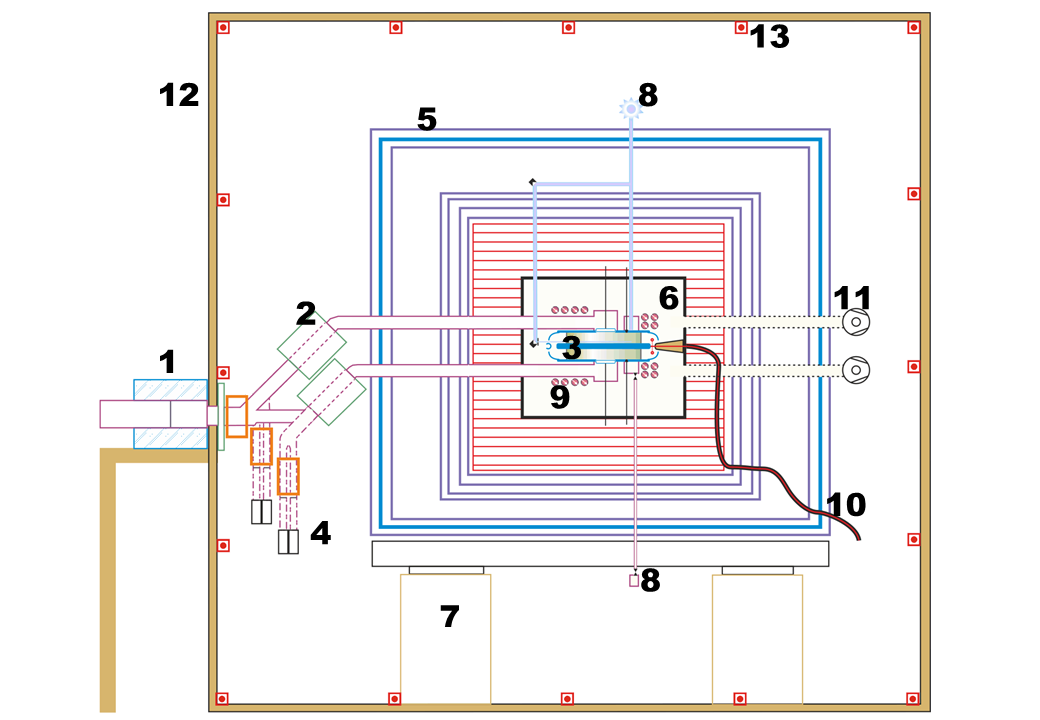}
  \caption{\textbf{Experimental apparatus.} The UCN coming from the source are polarized after passing a 5~T superconducting polarizer magnet (1). \corrcol{Two switches (2), containing each 2 UCN guides are used to fill and empty the UCN chambers (3).} After a typical precession time of 180~s, the UCN are counted in the detectors (4). The storage chambers and the vacuum vessel (6) are in a magnetically shielded room (5), which rests on an Aluminum frame supported by four granite pillars (7). The magnetic field is monitored in situ with a Hg system (8) and a Cs magnetometer array (9). The high voltage is provided via a cable (10) and the vacuum vessel is pumped with turbo molecular pumps situated outside of the magnetically shielded room (11). The entire setup is inside an insulation shell, thermally stabilized by air-conditioning (12). A surrounding field compensation (SFC) system will actively reduce the magnetic perturbations of the environment (13).}
\label{fig:shield}
\end{figure}
\begin{figure}[h]
\centering
  \includegraphics[width=0.5\textwidth]{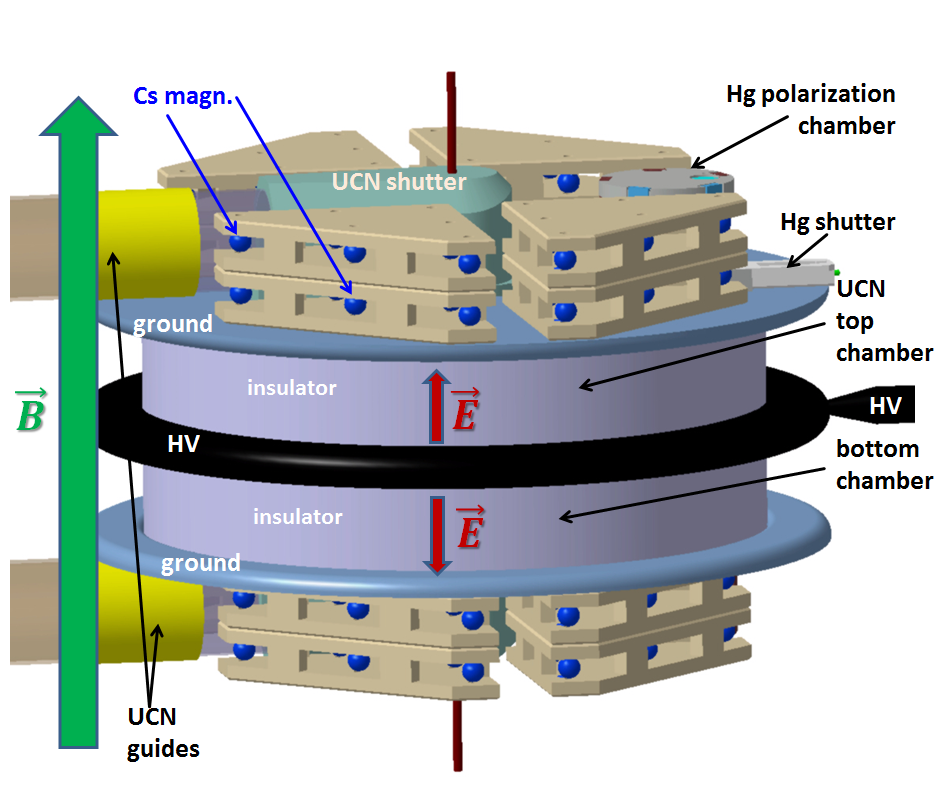}
  \caption{\textbf{Double chamber design.} The center electrode is at high voltage while the two outer electrodes are grounded. Both UCN chambers will also contain polarized Hg atoms used as a co-magnetometer. An array of Cs magnetometers will surround the top and bottom electrodes. This entire setup will be placed in a uniform magnetic field of 1~$\mu$T.}
\label{fig:electrodes}
\end{figure}

The n2EDM apparatus will significantly improve the neutron counting statistics and lower systematics with respect to its predecessor. 
A factor of 10 improvement in sensitivity compared to the present best limit \cite{pendlebury2015, baker2006} is calculated, based on the known performance of the \corrcol{previous} apparatus, \corrcol{the average UCN source performance in 2016 and a full simulation of the n2EDM apparatus benchmarked with several measurements.}
%
%
%
The experiment will profit from the high UCN intensity provided by the PSI UCN source~\cite{ries2017, lauss2014}.
The experimental apparatus is shown in Fig.~\ref{fig:shield}.\\
%
%
%
%
%
%
The experimental setup is based on our experience with the previous nEDM apparatus, the main difference being a double chamber design as pioneered in~\cite{altarev1980}. 
This double chamber design, as shown in Fig.~\ref{fig:electrodes}, has a number of advantages including a direct increase of the statistics through a larger total UCN volume with a diameter of 80 cm.
%
%
The high voltage electrode is at the center of the stack with the two ground electrodes connected to the rest of the setup.
This way, E-fields up to 15 kV/cm are expected.
Most importantly, we will be able to measure in both field configurations of eqs. \ref{eq:upup} and \ref{eq:updown} simultaneously.
This will strongly reduce any time dependent systematic effects.\\
%
%
%
%
%
%
%
%
The following sections will describe the subsystems of n2EDM as they are planned for the baseline setup.
%
%
%
\subsection{Magnetically shielded room}
The new magnetically shielded room\footnote{VAC GmbH, Hanau, Germany (https://www.vacuumschmelze.com)} (MSR) for n2EDM will comprise of 6 cubic layers of mu-metal and one additional layer of aluminum for RF-shielding, see Fig.~\ref{fig:shield}.
The MSR consists of an outer and inner cabin where enough space is left between the cabins for an intermediate room.
%
This room will contain the equipment sensitive to electro-magnetic noise (e.g., pre-amplifiers, current sources, \corrcol{etc.}).
The innermost shielding layer is a cube of approx.\ 3~m side length.
The innermost room \corrcol{with its 2 m $\times$ 2 m door} is large enough for a vacuum tank with approximate dimensions 1.9 m $\times$ 1.9 m $\times$ 1.5 m.
\corrcol{The MSR has 3 doors on one side and more than 70 openings in the walls to operate the experiment.}
%
%
The largest have a diameter of 220~mm, adapted to the planned UCN guide diameter.
Similarly sized openings are placed on the opposite side of the shield for field symmetry and uniformity reasons.
These will be used for the main pumping tubes of the vacuum vessel.\\
In order to achieve the planned statistical and systematic sensitivity, the MSR needs to provide a magnetically stable and uniform environment.
%
%
For the magnetic shield, a quasistatic shielding factor better than 80’000 at 0.01 Hz is expected.
The residual magnetic field is expected to be less than 0.5~nT \corrcol{over the volume covering the two precession chambers.}
The gradient at that location should be smaller than 3~pT/cm.
Each layer of the MSR is equipped with a separate set of degaussing coils.
The ability to degauss each layer in this manner helps to obtain uniform residual magnetic fields~\cite{voigt2013measures}.
%
%
\subsection{Surrounding field compensation}
The n2EDM experiment is located in the vicinity of other facilities generating strong magnetic fields (e.g., SULTAN, COMET)~\cite{schippers2007sc,bruzzone2008test}. 
They induce changes in the magnetic field at the experiment location of up to tens of $\mu$T on time scales from minutes to hours~\cite{afach2014}. 
Even though the n2EDM measurement volume will be shielded from these perturbations by the MSR, we might be left with measurable changes in the magnetic field of the experiment.
An active surrounding field compensation (SFC) system is being developed as an additional shielding layer and might be installed after initial characterization measurements.\\
A new method for designing coils of arbitrary field shapes, to compensate specific gradients and fields was developed~\cite{rawlik2018simple}.
%
%
In our case, the coils will be mounted on a grid which will surround the MSR.
This grid will be able to support a large number of sensors and coils which generate complex fields and gradients.
Additionally, coils tailored for specific magnetic field disturbances are considered as an option.
This way, specific magnetic field changes produced by experiments could be compensated.
%
%
\subsection{Magnetic field}
\begin{figure}[h]
	
	\begin{subfigure}{\textwidth}
		\includegraphics[width=0.45\textwidth]{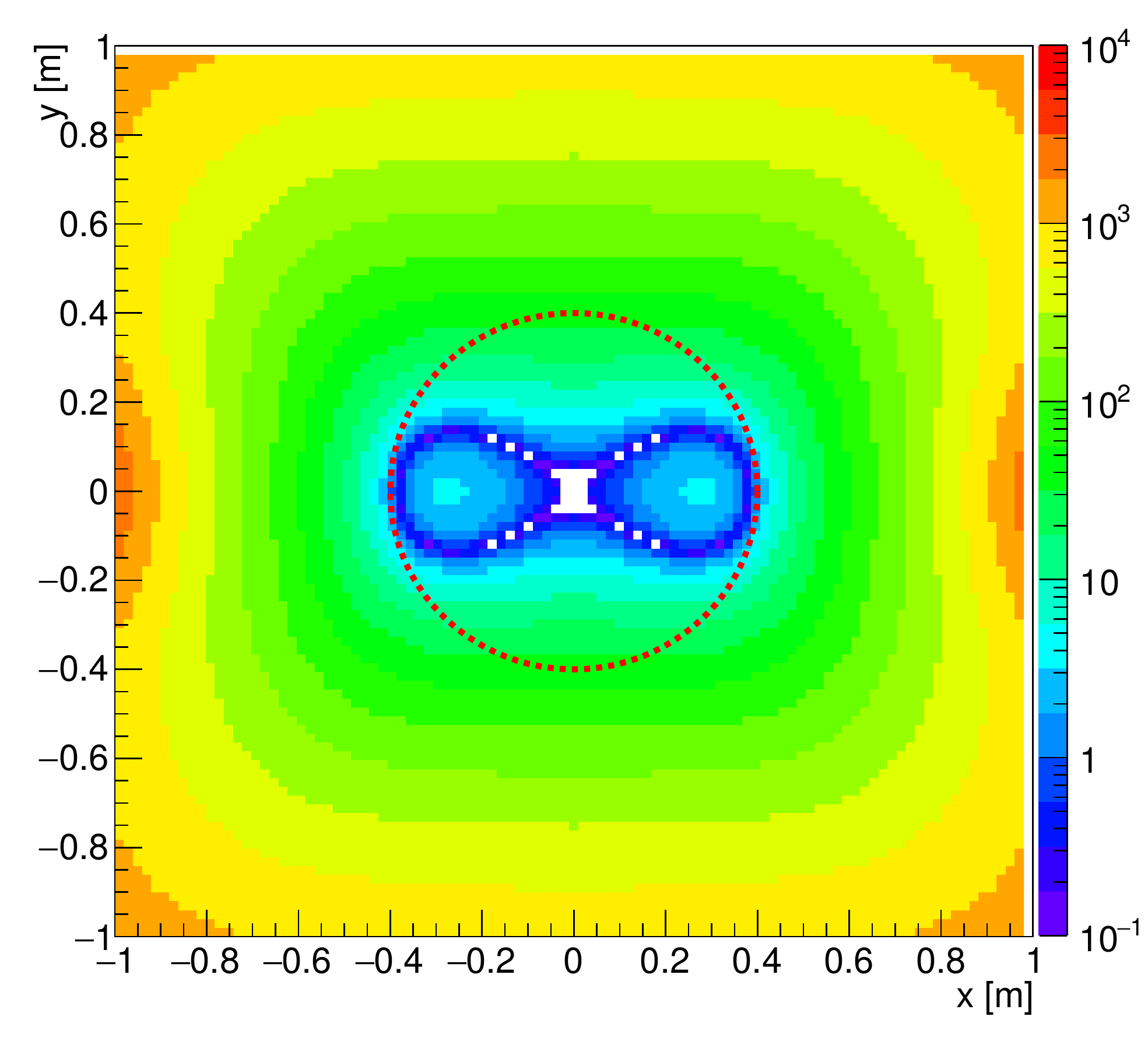}
		\caption{$B_0$ uniformity in the XY plane in pT.}
		\label{fig:coilH}
	\end{subfigure}
	
	\begin{subfigure}{\textwidth}
		\includegraphics[width=0.45\textwidth]{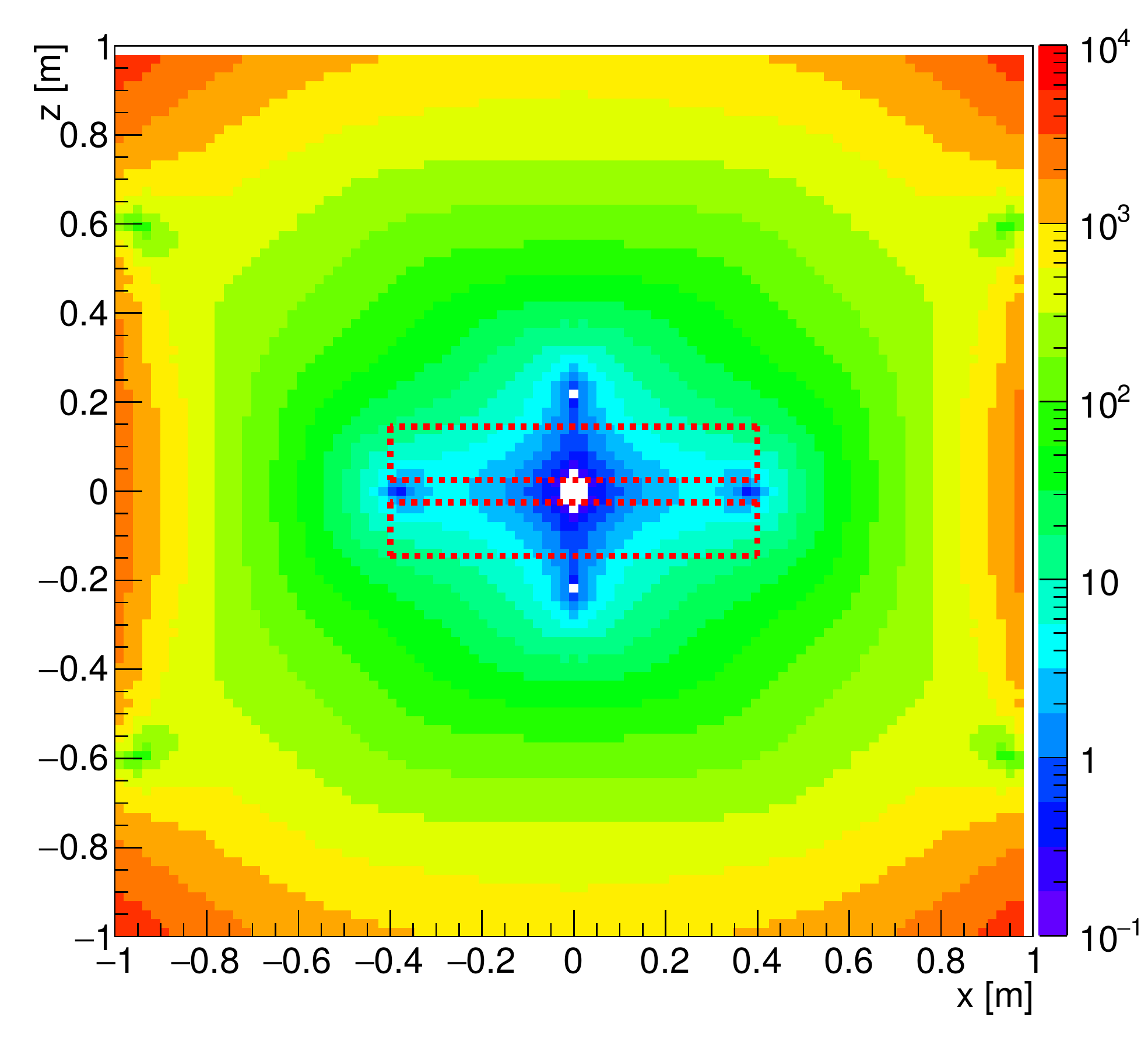}
		\caption{$B_0$ uniformity in the XZ plane in pT.}
		\label{fig:coilV}
	\end{subfigure}
	
	\caption{\textbf{$B_0$ field uniformity.} Calculated magnetic field deviation in pT from the nominal $1~\mu$T for the $B_0$ coil. The $B_0$ field is aligned with the z-axis on the graphs. The red disc (\ref{fig:coilH}) and the boxes (\ref{fig:coilV}) represent the UCN chambers with a diameter of 80~cm and a height of 12~cm. The UCN chambers are within the $10^{-4}$ field uniformity region, i.e., 100~pT on a $1~\mu$T field.}
	\label{fig:coil}
\end{figure}
n2EDM requires a very stable and uniform magnetic field, $B_0$.
The main reasons are to precisely control gradients which induce motional false EDMs and to ensure that the mean field is the same in both UCN chambers~\cite{afach2015measurement,pignol2018}.
%
%
A set of dedicated coils inside the innermost shielding layer will generate the $B_0$ field.
The coils need to be supplied with a very low noise and stable current to yield the best possible performance.\\
The target for n2EDM is a magnetic field uniformity better than $10^{-4}$ in the UCN chambers.
%
%
The $B_0$ coil is a single layer cubic solenoid mounted inside the MSR, around 10 cm from the innermost mu-metal layer.
The calculated field uniformity for this coil is shown in Fig. \ref{fig:coil} for a $1~\mu$T field in absolute terms, as we are interested in the absolute accuracy.\\
%
%
%
%
%
Even though the field uniformity in the simulation already reaches our target value, perturbations in the magnetic environment will worsen this performance.
The target magnetic field uniformity will be achieved with a $B_0$ coil accompanied by a set of correcting trim coils, which will be installed on the same frame as the $B_0$ coil. 
An array of 56 \corrcol{rectangular} trim coils will be used to produce all generic field gradients up to the 6th order~\cite{rawlik2018simple}.\\
Additionally, we require a long term stability of the $1~\mu$T $B_0$ field on the order of 30 fT for a period of about 300 s, i.e., one measurement cycle.
This field stability is necessary \corrcol{in order to reduce the error contribution due to the Hg co-magnetometer below 2\% of the neutron statistical error~\cite{ban2018}}.
The longterm stability depends on the stability and the noise level of the current source supplying the $B_0$ coil.
Therefore, we are developing an atomic magnetic resonance based current controller which can actively stabilize the current in the $B_0$ coil~\cite{koss2017}.\\
%
%
\subsection{Magnetometry}
With the MSR, the SFC and the set of coils in the innermost shielding layer, we expect to achieve the required magnetic environment in the UCN chambers.
However, experience from previous nEDM experiments has shown that it is important to accurately monitor the magnetic field in the experimental volume~\cite{green1998}.
For this purpose we will use several different types of magnetometers inside of the vacuum vessel.\\
First, a mercury ($^{199}$Hg) co-magnetometer occupying the same volume as the UCN.
Second, an array of cesium ($^{133}$Cs) magnetometers which surrounds the UCN volumes and measures the distribution of the magnetic field in the vacuum vessel.
\subsubsection{Hg co-magnetometer}
\label{sec:Hgco}
Optically polarized Hg atoms occupying the same volume as the UCN will measure the magnetic field in the UCN chambers.
With the magnetic field readings of the co-magnetometer, one can correct shifts in the neutron Larmor frequency $\nu_{\rm n}$ \corrcol{due to B-field changes as given in eq.~\ref{eq:singleR}.}
%
%
The \corrcol{first} phase of the experiment used such a Hg co-magnetometer~\cite{baker2014}.
There, microwave-excited Hg lamps were used for the optical pumping and probing of the Hg medium.
Meanwhile, we have developed a laser-based Hg co-magnetometer.
The light for the optical pumping and probing of the Hg medium will be delivered by a single laser.
This \corrcol{new laser-based system} system has a 5.5 times higher signal-to-noise ratio than its lamp-based predecessor with a sensitivity of 5~fT~\cite{ban2018}.
\subsubsection{Cs magnetometer array}
The primary goal of the Cs magnetometer array is to provide the necessary information about the magnetic field uniformity in the experiment. 
The field uniformity is characterized by a set of gradient multipoles which will be estimated using magnetic field readings from many independent magnetometer modules~\cite{pignol2018,knowles2009}.
%
%
%
We developed a method to make the $B_0$-field more uniform using the readout of all Cs magnetometers and the correcting trim coils of the experiment~\cite{wursten2018csarray}.\\
%
%
The new Cs magnetometer array envisioned for the n2EDM experiment is shown in Fig.~\ref{fig:electrodes}.
The sensors will be of the Bell-Bloom type~\cite{blum1961}, where the Cs medium is polarized by modulating the pumping rate at the Larmor frequency.
More specifically, we intend to use AM-modulated pumping of the medium followed by a free spin precession (FSP) period.
The FSP operation mode yields a sensitivity < 100~$\rm fT/\sqrt{\rm Hz}$~\cite{grujic2015}.
Additional advantages of this mode are the all-optical nature of the sensor, i.e., being magnetically silent, with low systematics on the Larmor frequency readout.
Recent design studies suggest that a particular sensor arrangement in the array will enhance its performance.
%
\subsection{UCN system}
\begin{figure}[h]
	\centering
	\includegraphics[width=0.5\textwidth]{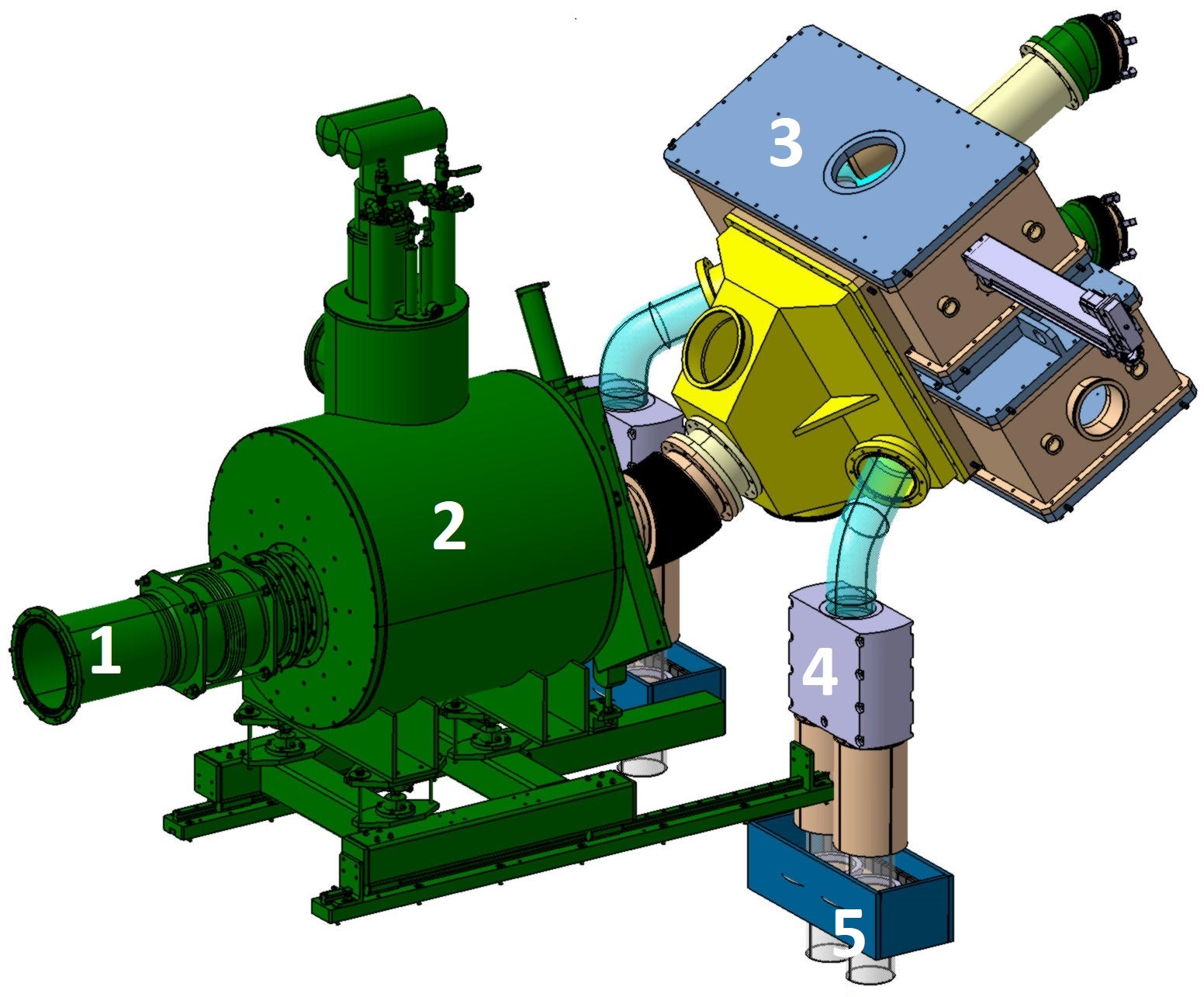}
	\caption{\textbf{UCN system.} The UCN are supplied from the UCN source (1). The UCN are polarized after passing a 5~T superconducting polarizer magnet (2). The polarized UCN are guided into the UCN chambers through a UCN switch (3). After the measurement cycle, spin analyzers (4) select \corrcol{separate} spin states of the UCN and guide them to \corrcol{separate neutron detectors for each spin state}. Only the detector mount points are shown at (5).}
	\label{fig:detection}
\end{figure}
%
%
%
UCN statistics are the main limiting factor for all nEDM experiments.
Therefore, the design of the n2EDM experiment was based on optimizing the UCN statistics for our UCN source.
The position of the UCN chambers and the guiding of the UCN from the UCN source to the UCN chambers \corrcol{were} optimized using the MCUCN code~\cite{zsigmond2018}.\\
We have developed guides which will be made from DURAN glass tubes with sub-nanometer surface roughness~\cite{bondar2017}.
Some bent parts will be machined from aluminum.
Non-magnetic nickel-molybdenum coating will be sputter-coated on the inside of the tubes and bends using the PSI coating facility.\\
The so-called ``switch'' is located between the superconducting polarizer magnet and the UCN chambers. 
Its function is to allow the selection of different guiding paths for the UCN.
First, the UCN are guided from the source to the UCN chambers during an initial filling period.
Then, after a precession time of typically 180~s, the UCN are guided to the UCN detectors.
The switch must work very precisely since all UCN guiding parts must be carefully aligned in order to minimize gaps and UCN losses.\\
%
%
%
%
%
%
%
Each UCN chamber will be emptied into a simultaneous spin analyzer, which is able to count neutrons for both spin components at the same time.
A similar system was already employed in the \corrcol{previous} apparatus~\cite{afach2015}.
Each arm of the spin analyzers will be equipped with a \corrcol{high-rate} neutron detector.
%
%
%
%
While the \corrcol{previous} spectrometer used $^6$Li-doped glass scintillators~\cite{ban2016}, we are investigating a gaseous scintillator operating at atmospheric pressure with a gas mixture of CF$_4$ and $^3$He for n2EDM.
%
%
This detector has the advantage of a faster response and lower background compared to the  $^6$Li scintillator.
\section{Expected statistical sensitivity}
The apparatus as described above represents the baseline setup for n2EDM.
We have performed a simulation of the full apparatus connected to the UCN source using MCUCN to determine the statistical sensitivity of \corrcol{n2EDM}, see Table~\ref{tab:sensitivity}.
This simulation was based on the average UCN source performance in 2016 \corrcol{and on absolute rate calibration measurements on the PSI West-1 beamline.}
The improvement in UCN statistics mostly comes from the two much larger UCN chambers and \corrcol{from the UCN guides with improved coating and larger diameter which are adapted to the PSI UCN source.}
\corrcol{The vertical positioning of the chamber was optimized with respect to the UCN energy spectrum and the material optical potential of the precession chamber coating material.}
%
%
%
It is clear that any improvement of the UCN source will immediately improve the statistical sensitivity of the n2EDM experiment.\\
The UCN storage time is limited by the quality of the coating and the gaps of the UCN chambers.
For the estimate of the expected sensitivity we \corrcol{keep the previously accomplished} performance with a free precession period $T$ = 180~s.\\
The electric field E was limited to 11kV/cm in the predecessor experiment by the presence of many optical fibers leading to the HV electrode, used to operate the Cs magnetometer array~\cite{knowles2009}.
%
%
%
Without these fibers, stable operation up to 15 kV/cm was achieved.
The double chamber design of n2EDM will remove these difficulties since all Cs magnetometers will be grounded.
Thus, operation at 15 kV/cm is expected.\\
The visibility, $\alpha$, is a measure of the UCN polarization after storage in the UCN chambers.
The best initial polarization achieved in the \corrcol{previous} apparatus was $\alpha_0 = 0.85$ while the average was $\alpha = 0.75$~\cite{wursten2018csarray}.
%
%
Three different depolarization mechanisms will decrease the UCN polarization.
First, the depolarization of UCN induced by wall collisions.
%
%
Second, the intrinsic depolarization where all UCN are affected by a magnetic gradient.
 Last, the gravitationally enhanced depolarization where UCN of different energy classes acquire phases at different rates~\cite{afach2015stri,afach2015gravitational}.
We expect the uniformity of the $B_0$ field to be sufficient to obtain an average polarization of $\alpha = 0.8$.
%
%
%
%
\begin{table}[h]
\centering
\begin{tabular}{l|c|c}
   & nEDM 2016     & n2EDM baseline \\
\hline \hline
diameter (cm)      & 47  & 80 \\
\hline
$\alpha$           & 0.75 & 0.8 \\
$E$ (kV/cm)        & 11   & 15  \\
$T$ (s)            & 180  & 180 \\
$N$ (per cycle)    & 15'000 & 121'000 \\
\hline
\rule{0pt}{3ex}    
$\sigma (d_{\rm n})$ (per day) & $11 \times 10^{-26}$ \ecm & $2.6 \times 10^{-26}$ \ecm \\
\rule{0pt}{3ex}
$\sigma (d_{\rm n})$ (total) & $9.8 \times 10^{-27}$ \ecm & $1.1 \times 10^{-27}$ \ecm \\
\hline \hline
\end{tabular}
\caption{\textbf{Statistical sensitivity \corrcol{at 68\% C.L.}} Comparison between the achieved performance of the \corrcol{previous} apparatus and the estimate for n2EDM as described in the text. The total sensitivity estimate for nEDM 2016 is based on the total data acquired with the predecessor experiment until the end of 2016. The total sensitivity estimate for n2EDM is based on projected 500 days of measurement \corrcol{with about 280 cycles per day. PSI usually operates the proton beam and the UCN source on about 200 days per calendar year.}}
\label{tab:sensitivity}
\end{table}
\section{Systematics}
In reference~\cite{pendlebury2015} we have laid out a large set of systematic effects which have to be dealt with during the analysis of our acquired nEDM data.
%
%
%
The systematic effects can be classified in two different categories: direct and indirect.
A direct systematic effect means that a variation of a measurement parameter immediately affects the value of the extracted nEDM.
An indirect systematic effect is introduced through data analysis~\cite{pendlebury2004}.\\
%
%
%
%
%
Here we shall describe two important systematic effects of the \corrcol{previous} apparatus and how they are intended to be controlled in n2EDM.
\subsection{Uncompensated E/B-field correlation}
The application of the electric field might itself generate a change in the magnetic field which is correlated with the electric polarity.
%
%
This is a major concern in any EDM experiment as it can produce a direct systematic effect. 
Such an effect might be due to the leakage current from the high voltage electrode to the ground electrodes. 
It could also be due to magnetization of a part of the apparatus by charging currents during voltage ramps.\\
In principle, the correction using the mercury co-magnetometers cancels any magnetic field fluctuations, including those correlated with the electric field. 
However, the cancellation is not perfect due to the gravitational shift~\cite{afach2015stri}.
This shift arises due to the difference in center-of-mass between the Hg atoms and the UCN, i.e., $\langle z \rangle$.
The false EDM due to a correlated part of the gradient $\delta G(E)$ for a double chamber design reads
\begin{equation}
d^{\rm false}_{\rm n} = \frac{\hbar | \gamma_{\rm n} |}{4E} (\langle z \rangle ^{\rm B} - \langle z \rangle ^{\rm T}) \delta G (E)~.
\end{equation}
In the double chamber design only the difference in the center-of-mass offsets contributes.
Therefore the effect will be strongly reduced.
%
%
The goal for n2EDM is to have this systematic effect under control at the level of $5\times 10^{-28}$ \ecm.
Assuming E = 15 kV/cm, and $\langle z \rangle ^{\rm B} - \langle z \rangle ^{\rm T}$ = 0.1 cm, this corresponds to a control over the correlated part of the gradient at the level of G(E) $\leq$ 2~fT/cm.
The Cs magnetometer array is intended to control this effect.
%
%
%
%
%
%
%
\subsection{Motional false EDM}
%
%
When a particle moves with a velocity, $\vec{v}$, with respect to a static electric field, $\vec{E}$, it \corrcol{is affected by} a motional magnetic field $\vec{B}_{\rm m} = \vec{E} \times \vec{v}/c^2$. 
In our case, we consider the motion of the UCN and Hg atoms in the UCN chambers.
For the trapped particles the velocity averages to zero and therefore one is naively led to conclude that the effect vanishes. 
However, $\vec{B}_{\rm m}$ does induce a Larmor frequency shift linear to the electric field when the particles evolve in a non-uniform magnetic field~\cite{pignol2012,pendlebury2004}.
In these conditions, a given particle with the trajectory $\vec{r}(\tau)$ is subjected to a time dependent transverse magnetic field given by $B_x (\tau) = B_x(\vec{r}(\tau)) + B_{{\rm m},x}$ and $B_y (\tau) = B_y(\vec{r}(\tau)) + B_{{\rm m},y}$ , where $B_x$ and $B_y$ are the transverse components of the $B_0$ field.\\
Redfield’s theory provides a general approach to calculate frequency shifts and relaxation rates on a quantum system caused by a randomly fluctuating perturbation~\cite{pignol2012,goldman1988}.
Specifying this theory to the problem of spins in a bottle, one finds that the fluctuating transverse magnetic field seen by the
individual particles induces a frequency shift on the ensemble proportional to $E$.
This leads to a false EDM effect which is given by
\begin{equation}
d^{\rm false} = \frac{\hbar \gamma^2}{2c^2} \int^{\infty}_0 d\tau \cos (\omega \tau) \langle B_x(0) v_x (\tau) + B_y (0) v_y (\tau) \rangle~,
\label{eq:motEDM}
\end{equation}
where $\omega = \gamma_{\rm Hg} B_0$ and $\langle \cdot \rangle$ is the field-velocity correlation function of the particles in the UCN chamber.
The gyromagnetic ratio in eq. (\ref{eq:motEDM}) depends on the species considered, and can be both $\gamma_{\rm n}$ or $\gamma_{\rm Hg}$.\\
Because the mercury co-magnetometers will be used to correct the neutron frequency in each chamber for the fluctuations of the magnetic field, the false EDM on the mercury atoms will appear as a false neutron EDM, with a magnitude of
\begin{equation}
d^{\rm false}_{\rm Hg \rightarrow n} = \bigg| \frac{\gamma_{\rm n}}{\gamma_{\rm Hg}} \bigg| d^{\rm false}_{\rm Hg} = 2 \times 10^{-26} \ecmm \times G~({\rm pT/cm})~,
\end{equation}
\corrcol{where $G$ (pT/cm) is the linear gradient over the volume of the chambers.}
This induced false EDM is two orders of magnitude larger than the direct false EDM $d^{\rm false}_{\rm n}$.\\
The general strategy to cancel this effect is to split the data-production into many runs with different \corrcol{vertical} gradient configurations.
Then, the measured EDM is plotted as a function of the gradient and extrapolated to zero gradients.
If managing the effect through applying different gradients proves too difficult, we have devised a way to operate the n2EDM spectrometer at higher $B_0$ field strengths.
For certain UCN chamber diameters at higher $B_0$ field strengths, the integral in eq. (\ref{eq:motEDM}) vanishes and the false EDM effect becomes zero~\cite{pignol2018magic}.
\section{Conclusion}
We have presented the new n2EDM spectrometer for the nEDM search at PSI. 
The design of the apparatus is based on the technical expertise which our collaboration has gained during the first phase of the experiment. 
Its design is guided by optimizing neutron statistics under adequate control of corresponding systematics.
The large volume double chamber setup will allow us to increase the statistical precision of the measurement as well as to better control some systematic effects.
Its large size however makes the requirement on magnetic field control much more stringent.
The large size of the magnetically shielded room and of the magnetic field coils array will provide the necessary magnetic field stability and uniformity.
The two Hg co-magnetometers and the large array of high precision Cs magnetometers will provide the necessary field monitoring and control.
With this new setup and conservative performance estimates based on the previous apparatus, we expect to reach a sensitivity of 1$\times$10$^{-27}$ \ecm in 500 days of data taking using the described baseline setup.
Upgrades and changes to this apparatus are currently subject of intense research and could further expand the n2EDM sensitivity into the 10$^{-28}$ \ecm range.
%
%
%
%
%
%
%
%
\section*{Acknowledgments}
The experiment could not be realized without the many
excellent ideas and continuous dedicated work of 
D.~Goupillière (LPC), 
M.~Meier (PSI),
J.~Menu (LPSC), 
Y.~Merrer (LPC) and  
T.~Stapf (PSI).
We acknowledge the continuous outstanding support 
at LPC by
B.~Bougard, 
B.~Carniol, 
P.~Desrues, 
D.~Etasse, 
J.M.~Fontbonne, 
C.~Fontbonne, 
J.~Hommet,  
J.~Lory, 
C.~Pain, 
J.~Perronnel, 
J.~Poincheval,  
H.~de Préaumont, 
C.~Van Damme;\newline
at LPSC by
M.~Chala,
R.~Faure,
C.~Fourel,
J.~Fulachier,
C.~Geraci,
J.~Marpaud,
C.~Martin,
M.~Marton,
J.~Odier,
S.~Roni,
S.~Roudier,
J.P.~Scordilis,
C.~Thomassé,
C.~Vescovi; \newline
at PSI by
B.~Blau, 
K.~Boutellier,
F.~Burri,
P.~Erisman, 
A.~Ersin, 
A.~Gn\"adinger, 
U.~Greuter, 
J.~Hadobas, 
L.~Holitzner,
M.~Horisberger, 
B.~Jehle, 
R.~K\"ach, 
G.~K\"aslin, 
C.~Kramer, 
M.~M\"ahr, 
M.~M\"uller,
O.~Morath, 
W.~Pfister, 
D.~Reggiani, 
R.~Schwarz,
V.~Talanov, 
V.~Teufel, 
A.~van-Loon, 
\corrcol{X. Wang,}
J.~Welte,
M.~Wohlmuther; \newline
and at Sussex by
D.~Shires.\\
We are grateful to many PSI support groups.\\
P. Mohanmurthy acknowledges grant SERI-FCS 2015.0594.
STFC, via grants ST/M003426/1, ST/N504452/1 and ST/N000307/1; the School of Mathematical and Physical Sciences at the University of Sussex for a studentship and other financial support.
Support by the Swiss National Science Foundation Projects 200020-137664 (PSI), 200021-117696 (PSI), 200020-144473 (PSI), 200021-126562 (PSI), 200021-181996 (Bern), 200020-172639 (ETH) and 200020-140421 (Fribourg) is gratefully acknowledged. 
The LPC Caen and the LPSC Grenoble acknowledge the support of the French Agence Nationale de la Recherche (ANR) under reference ANR-14-CE33-0007 and the ERC project 716651-NEDM.
The Polish collaborators wish to acknowledge support from the National Science Center, Poland, under grant no. 2015/18/M/ST2/00056.
This work was partly supported by the Fund for Scientific Research Flanders (FWO), and Project GOA/2010/10 of the KU Leuven. 
We also greatly acknowledge granting access to the computing grid infrastructure PL-Grid\footnote{PLGrid-Consortium, Polish Grid Infrastructure PL-Grid, 2017. http://www.plgrid.pl/en}.
%
%
\bibliography{n2EDM_proc}

\begin{thebibliography}{46}
\bibitem{pignol2018magic}
G.~Pignol, ``A magic magnetic field to measure the neutron electric dipole
  moment'', To be published
\end{thebibliography}
%
%
%

\end{document}